\pdfoutput=1 
\documentclass[journal]{IEEEtran}

\usepackage{multirow}
\usepackage{cite}
\usepackage{hhline}
\usepackage{graphicx}
\usepackage{subfigure}
\usepackage{setspace}
\usepackage{stfloats}
\usepackage{color}
\usepackage{cite}
\usepackage{amsmath, amsfonts, amssymb}
\usepackage{epstopdf}
\graphicspath{{./fig/}}

\begin{document}
%
\title{Sum-Rate Maximization and Data Delivery for Wireless Seismic Acquisition}

\author{\IEEEauthorblockN{First Last}
\IEEEauthorblockA{School of Electrical and\\Computer Engineering\\
University of Abcdefg\\
Email: email@domain.cedu}
\and
\IEEEauthorblockN{First Last}
\IEEEauthorblockA{School of Electrical and\\Computer Engineering\\
	University of Abcdefg\\
	Email: email@domain.cedu}
\and
\IEEEauthorblockN{First Last}
\IEEEauthorblockA{School of Electrical and\\Computer Engineering\\
	University of Abcdefg\\
	Email: email@domain.cedu}}


%
\author{\IEEEauthorblockN{Abdullah Othman\IEEEauthorrefmark{1}\IEEEauthorrefmark{3},
Wessam Mesbah\IEEEauthorrefmark{1}\IEEEauthorrefmark{3},
Naveed Iqbal\IEEEauthorrefmark{1}\IEEEauthorrefmark{3},
Suhail Al-Dharrab\IEEEauthorrefmark{1}\IEEEauthorrefmark{3},
\\Ali Muqaibel\IEEEauthorrefmark{1}\IEEEauthorrefmark{3}},
Gordon St\"uber\IEEEauthorrefmark{2}\IEEEauthorrefmark{3}\\
\IEEEauthorblockA{\IEEEauthorrefmark{1}Electrical Engineering Department,
King Fahd University of Petroleum and Minerals, Dhahran, Saudi Arabia
\IEEEauthorblockA{\IEEEauthorrefmark{2}School of Electrical \& Computer Engineering, Georgia Institute of Technology, USA}}\\
\IEEEauthorblockA{\IEEEauthorrefmark{3}Center for Energy and Geo Processing}}


\maketitle

\begin{abstract}
Traditional seismic acquisition systems suffer from a number of difficulties related to telemetry cables that are used as a means of data transmission. Transforming the traditional seismic acquisition system to a wireless system has been considered as a potential solution to most of these difficulties. The wireless seismic acquisition system has to serve a huge aggregate data rate requirement as is usually the case in a large wireless sensor network. This paper considers the wireless acquisition system, and studies the maximum achievable transmission data rates from the geophones to the wireless gateways. Successive interference cancellation decoding is assumed to be used at the gateway nodes. We consider the problem of sum-rate maximization by optimizing the decoding process at each gateway node. The optimization searches for optimal decoding set at each gateway, i.e. which group of geophones will be decoded at each gateway. Various integer programming algorithms are proposed for solving the maximization problem. These optimization algorithms are simulated and compared among each other, where it is shown that the ant system algorithm achieves the highest sum-rate with lower computational complexity compared to other algorithms. Furthermore, the data delivery from the gateways to the data center is also considered. In this stage, two gateways with different buffer sizes are studied. For small-size buffers, two optimization problems are identified  and solved. The first problem considers the minimization of the total power of the gateways, and the second problem considers power fairness between the gateways. For large-size buffers, the problem of maximizing the weighted sum rate of the gateways is solved.
\end{abstract}


%
\IEEEpeerreviewmaketitle

\section{Introduction}
\begin{figure}
	\centering
	\includegraphics[width=7cm]{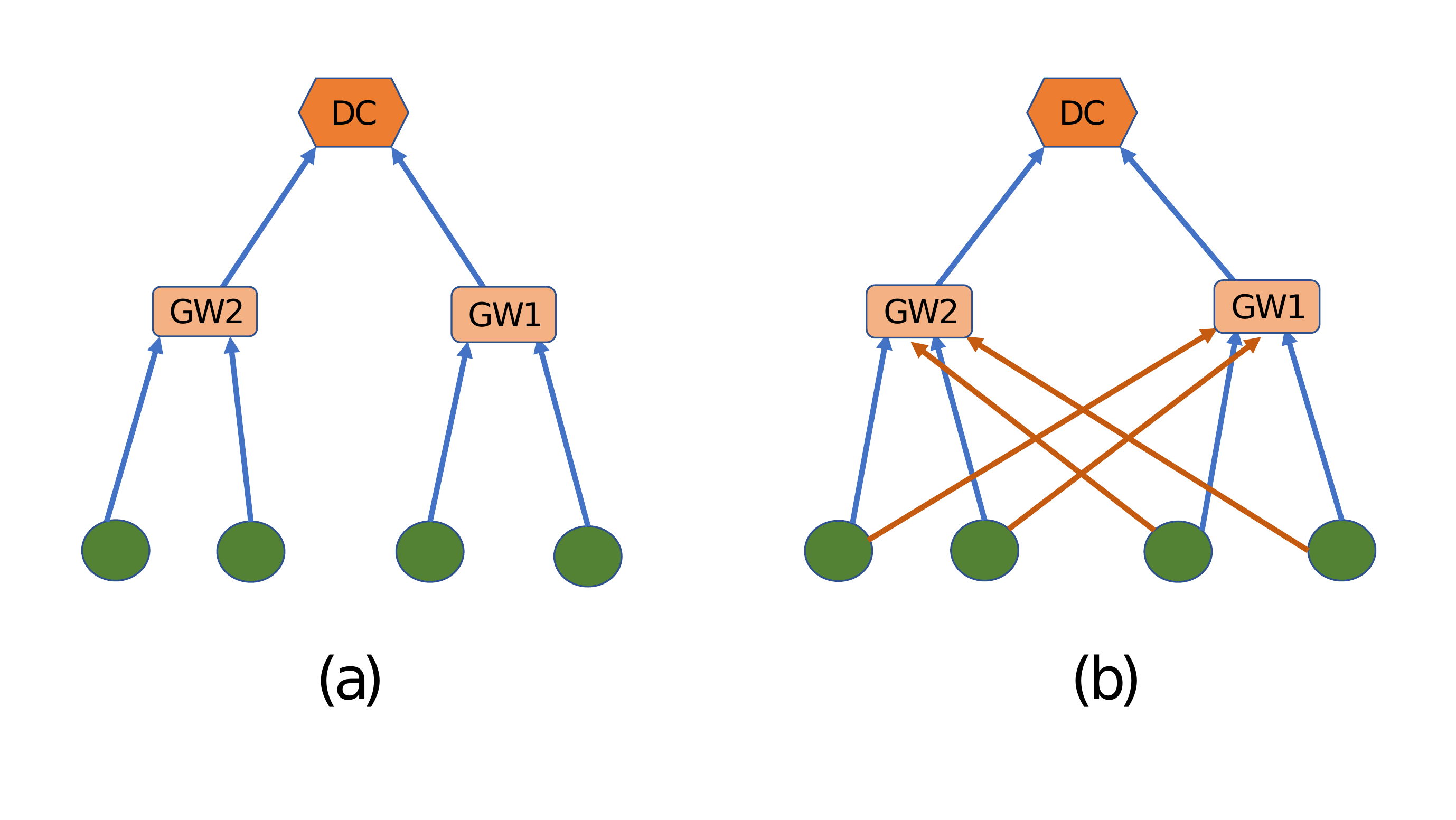}\\
	\centering
	\caption{a) A traditional tree structure and b) the modified tree adopted.}
	\label{fig:system_illustrated}
\end{figure}

Traditional seismic acquisition systems employ cables to efficiently transmit data from the geophones (GPs) to the data center (DC). Telemetry cables, although efficient and reliable, are a burden in terms of cost and weight. With surveys growing larger in scale, costs of deployment and maintenance become proportionally higher. Furthermore, environments with complex terrains render deployment of cabled systems impractical. Hence, there is an increasing trend in oil exploration companies to shift to wireless technology in order to avoid these difficulties.
Aspects for the transition to wireless acquisition include proper choice of wireless technology, the need for efficient communication protocols and reliable data transfer. Driven by the trend towards wireless seismic surveying, several authors studied cable-less surveys and wireless architectures in the literature. Savazzi \textit{et al.} \cite{Savazzi2013} proposed to use ultra-wide band wireless technologies in land seismic acquisition systems, where a hierarchical architecture-based, short-range and long-range communication were developed for two stage data delivery. Freed \cite{Freed2008} argued for nodal systems, where geophones receive seismic data continuously and store them for future retrieval. This eliminates the need for real-time data acquisition especially with a substantial number of channels. Ellis \cite{Ellis2014} proposed a hybrid system that combines cable and cable-less acquisition systems. Reddy \textit{et al.} \cite{Reddy2018} proposed a network architecture based on the IEEE 802.11af standard. The proposed architecture offers a high transmission data rate by operating in the TV white space. Most common wireless sensor network (WSN) topologies are tree, mesh and star topologies \cite{Hodge2015,Yang2015,Park2018,Sharma2013,Environments2013}. In a tree topology, the nodes are organized in a hierarchical structure. The root node serves intermediate nodes, that in turn serve other lower-level nodes. In a star topology, the central node is connected to all the other nodes. On the other hand, in a mesh topology, all the nodes are connected to each other through one-hop or multi-hop distances. The structure adopted in this work is an enhanced version of the tree structure where the gateways (GWs) act as intermediate nodes between the GPs and the DC. In contrast to the conventional tree topology, any GP is assumed to be capable of transmitting its data to all GWs, rather than to a single GW only (as depicted in Fig.~\ref{fig:system_illustrated}). This relaxes the constraints on the sum-rate achieved by the GPs, which is essential in the formulation of the first problem related to sum-rate maximization. To the best of our knowledge, this structure and the objective, i.e. sum-rate maximization of the GPs have not been studied in a previous work. Allowing the GPs to send their data to all the GWs encourages random placement of the GWs. Hence, drones can be used as GWs to receive and store GPs' data. This is an improvement compared to the work in \cite{Iqbal2018}, where GWs are assumed to be in fixed position.
Furthermore, in contrast to the previous works, and building on the work done in \cite{doi:10.1190/segam2018-2998141.1}, the focus of this paper is on the theoretical aspect of the geo-seismic network, which was found to be underrepresented in the literature. Study of upper bounds on the network sum-rate provides us with knowledge on the transmission capabilities of the geo-seismic network. It also presents a valuable insight on how the decoding and selection processes at the GWs could affect the achievable individual rates of the GPs as well as the combined sum-rate. For this purpose, comparisons between the case of optimizing the set of GPs to be decoded at each GW along with their decoding order, in order to maximize the achievable sum-rate and the case of no-optimization are presented.
In this work, the wireless system capability in terms of sum-rate is optimized. Information-theoretic bounds are used to gain insight into the achievable rates of the system. For this purpose, successive interference cancellation (SIC) is assumed at the (GWs). SIC is performed by successively decoding one of the GPs signals and subtracting it from the original signal, and then repeating the process for the rest of GPs, and thus reducing the interference after each decoding process. It is known that SIC achieves the capacity of the multiple access channel at each GW \cite{ElGamal2011}.

Furthermore, in this paper, the data transmission from the wireless GWs to the DC is considered. The DC uses SIC to decode the signals of the GWs and achieves the capacity of the multiple access channel (MAC). Although the problem in this work is formulated for seismic acquisition, it can be applied in general to other WSNs by taking the specifications into consideration.
In summary, the contributions of the paper are as follows:
\begin{enumerate}
\item The problem of maximizing the sum-rate of the GPs using SIC is formulated and three main metaheuristic algorithms are proposed to solve it.
\item The problem of data transmission from the GWs to the DC, for GWs with small buffers, is considered for two different objective functions and solved.
\item The problem of maximizing the weighted sum of the GWs rates is solved.
\end{enumerate}

The rest of the paper is organized as follows. In Section II, the system model is presented. In Section III, the mathematical formulation of the proposed optimization problems is discussed. The proposed integer programming algorithms to solve the first optimization problem are presented in Section IV and in Section V findings and discussions are shown. Finally, in Section VI the concluding remarks are presented followed by an appendix.

\section{System Model and Problem Description}
\begin{figure}
	\centering
	\includegraphics[width=9cm]{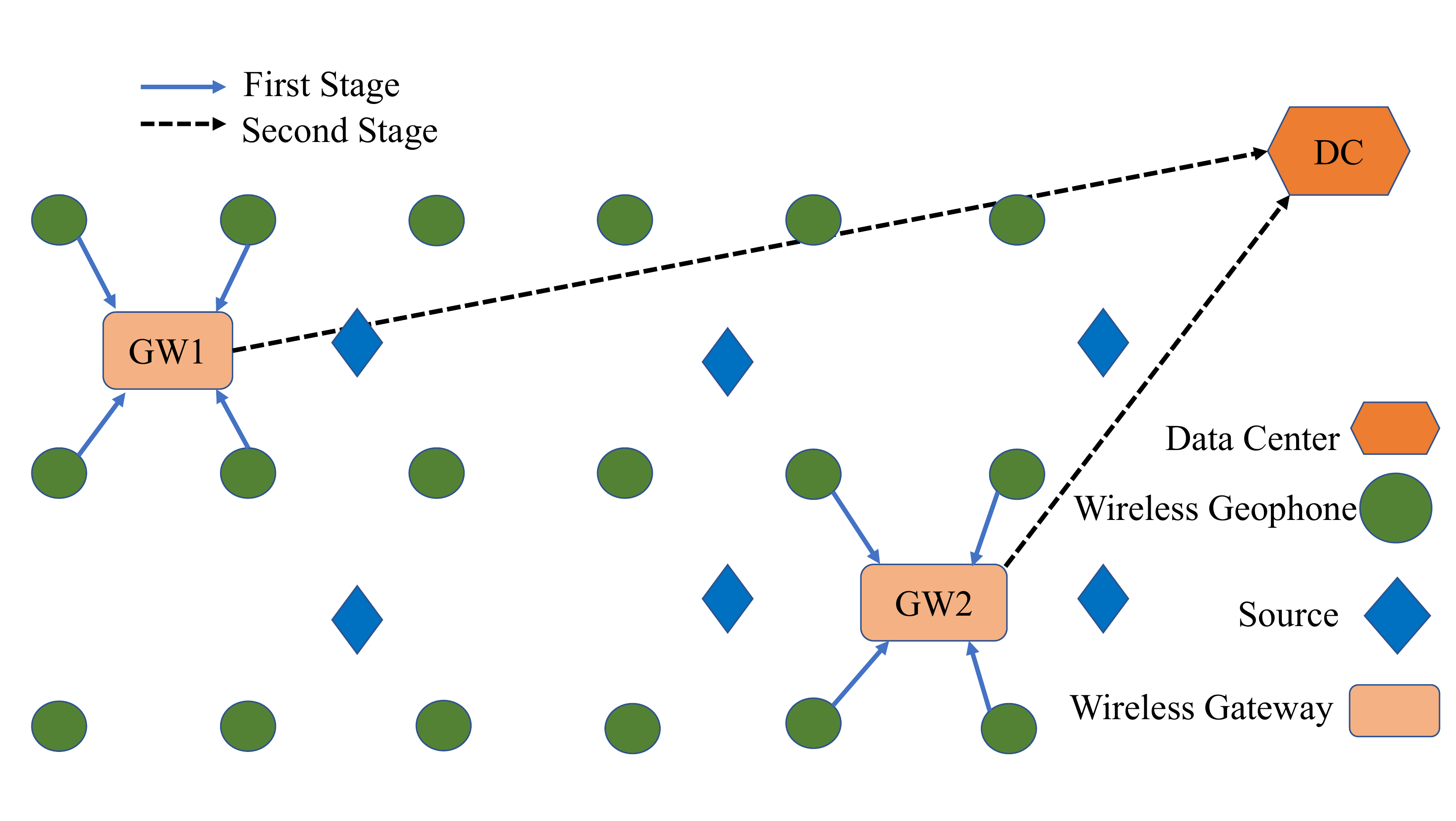}\\
	\centering
	\caption{A plane view of an orthogonal geometric wireless acquisition system}
	\label{fig:system model}
\end{figure}

In this work, the deployment of wireless GPs in a grid-like network to cover the whole survey area is considered, where wireless GWs are deployed between the GP lines. The purpose of GWs is to collect data from the GPs, possibly store them for some time, and then forward the collected data to the DC for processing \cite{Vermeer2012}. A typical land survey deploys 20,000 to 30,000 GPs over a large area on the order of 20 km\textsuperscript{2}. A GP equipped with a 24-bit analog-to-digital converter (ADC) generates sampled data at a rate of 144 kbps, with a sampling interval of 0.5 ms. The aggregate data rate for a typical survey can be about 4.32 Gbps, when all the GPs are active. A plane view of an orthogonal geometry survey is shown in Fig.~\ref{fig:system model}, where two GWs are placed between the source and receiver lines to harvest the data from the GPs. Fig.~\ref{fig:system model} also depicts the first and second stages considered in this work, where the first stage is the data transmission from the GPs to the GWs and the second stage is the data delivery from the GWs to the DC. The GWs may have some buffering capability to store the collected data, then it is sent to the DC for processing.

\begin{figure}
	\centering
	\includegraphics[width=7.5cm]{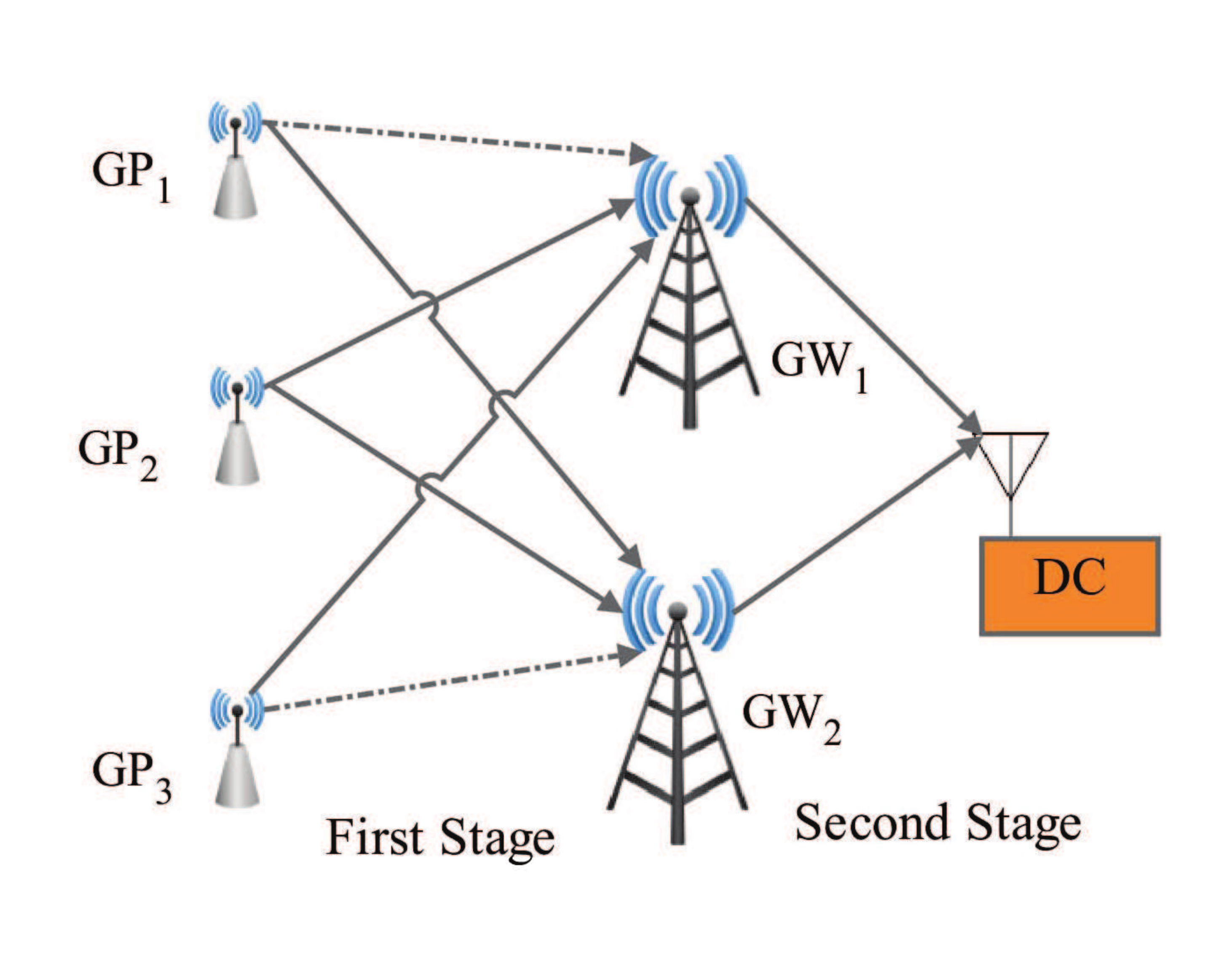}\\
	\caption{The two stages of seismic acquisition considered, for a small network of $K=3$  GPs and $N=2$  GWs}
	\label{fig:network example}
\end{figure}

In this paper, a system that consists of $K$  GPs and $N$  GWs is considered. The channels between the GPs and the GWs are assumed to follow a Rayleigh block fading model. This model is convenient for desert environment and seismic acquisition applications, where the GPs are placed in fixed positions and the scatterers are stationary. The channel matrix $H$ has $K\times N$  dimensions. Some assumptions about the system model are listed below:
\begin{itemize}
\item The channels between all the transmitters and receivers are constant for a fixed interval of time and change independently between time intervals.
\item The transmission period is a fixed time slot which is of equal width for all transmission sessions.
\item The channel state information (CSI) of all channels is available at all the GWs.
\item Each GP transmits with a fixed power, $P$.
\end{itemize}
The GWs are using SIC, where they can choose to decode some of the GPs and consider others as interference to maximize the total sum-rate of the network. Due to the use of SIC at the GWs, each GP will have different constraints on its rate, coming from the different GWs that decode its signal. The rate of each GP, $R_j$,  must satisfy all these constraints, which might lead to reducing its rate. Therefore, sometimes it is better not to decode the GP data at some GW as this will relax some of the constraints and help increase its rate. The Shannon normalized capacity is given as,
\begin{equation}
C=\log_2\Big (1+\frac{P|h|^2}{N_0} \Big ),
\end{equation}
where $C$  is in bits per second per Hz (bps/Hz), $h$  is the channel gain value taken from the channel matrix $H$. The additive noise is assumed to be Gaussian distributed with a variance $N_0$. If interference is present, it will appear in the denominator added to $N_0$. Each GP in a multiple access channel can transmit with a rate $R\leq C$. For multiple GPs transmitting simultaneously, a multiple access region is characterized by all the upper bounds on the rates.
It is assumed that each GW will decode the GP signals in a decreasing order, starting from the signal with highest signal-to-noise ratio (SNR) down to the signal with the minimum SNR. This will feature fairness to the weaker signals, while providing the maximum sum-rate \cite{ElGamal2011}. To illustrate how decoding selection at the GWs can affect the total sum-rate, consider a small representative network of $K=3$  GPs and $N=2$  GWs, $P=1\text{mW}$ , $N_0=1\text{mW}$  as shown in Fig.~\ref{fig:network example}. The channel gain matrix $H$  is generated following Rayleigh random density function and is given as,
\[H =
\left [
\begin{tabular}{ccc}
3.023 & 1.133 \\
1.738 & 2.168 \\
0.542 & 0.896
\end{tabular}
\right ].
\]
In this case, by going through all possible decoding order combinations, the one that gives the maximum sum-rate (denoted as max-sum-rate) is illustrated in the first stage in Fig.~\ref{fig:network example}, where GW\textsubscript{1} decodes GP\textsubscript{2} then GP\textsubscript{1} (solid arrow), while treating GP\textsubscript{3} as interference (dashed arrow), and GW\textsubscript{2} decodes GP\textsubscript{2} then GP\textsubscript{3}, while treating GP\textsubscript{1} as interference. To see the rate expressions for both cases please refer to Table \ref{max-sum-rate} and Table \ref{decode-all}. The rate expressions can be verified from (2) by considering the individual constraints on the rates. By substituting the values, the total sum-rate for the max-sum-rate case is found to be $\sum R \leq 3.813$ bps/Hz. Now compare this with the decode-all case, where all the GWs will decode all the GPs without doing any kind of optimization. In this case, it can be found that $\sum R  \leq 2.483$ bps/Hz. The individual rates for the two cases are shown in Table \ref{Casestudy1}. As can be seen, decoding a GP at multiple GWs puts more constraints on the rate of that GP, which might result in reducing its rate. However, this could present opportunities to increase the rates of other GPs, as is the case in max-sum-rate where the rate of GP\textsubscript{2} was decreased to 0.3668 bps/Hz, but it helped to increase the rate of GP\textsubscript{1} to 3.011 bps/Hz. Therefore, for some GPs, it might be better not to decode them at all GWs, while for others it will improve the sum-rate if they are decoded at all GWs. And this is the optimization problem that is considered in the first stage presented in the paper, i.e. how to optimize the set of GPs to be decoded at each GW so as to maximize the network sum-rate.
\begin{table}[]
\centering
	\caption{Rate Expressions for the max-sum-rate case}
	\label{max-sum-rate}
	\footnotesize{%
\begin{tabular}{|c|c|}
\hline
\hspace{-0.1cm}\textbf{GP}\hspace{-0.1cm} & \textbf{Max-sum-rate} \\ \hline
\hspace{-0.2cm}1 \hspace{-0.2cm} & $\log_2\big(1+\frac{P|h_{11}|^2}{N_0+P|h_{31}|^2}\big)$      \\ \hline
\hspace{-0.2cm}2 \hspace{-0.2cm} & \scriptsize{ $\hspace{-0.1cm}\min\hspace{-0.1cm}{\big[\log_2\hspace{-0.1cm}\big(1\hspace{-0.1cm}+\hspace{-0.1cm}\frac{P|h_{21}|^2}{N_0+P|h_{11}|^2+P|h_{31}|^2}\big),\log_2\hspace{-0.1cm}\big(1\hspace{-0.1cm}+\hspace{-0.1cm}\frac{P|h_{22}|^2}{N_0+P|h_{12}|^2+P|h_{32}|^2}\big)\big]\hspace{-0.1cm}}$                               }\\ \hline
\hspace{-0.2cm}3 \hspace{-0.2cm} & $\log_2\big(1+\frac{P|h_{32}|^2}{N_0+P|h_{12}|^2}\big)$        \\ \hline
\end{tabular}}
\end{table}

\begin{table}[]
\centering
	\caption{Rate Expressions for the decode-all case}
	\label{decode-all}
	\footnotesize{%
\begin{tabular}{|c|c|}
\hline
\hspace{-0.1cm}\textbf{GP}\hspace{-0.1cm}  & \textbf{Decode-all} \\ \hline
\hspace{-0.1cm}1\hspace{-0.1cm}    &   \scriptsize{$\min{\big[\log_2\big(1+\frac{P|h_{11}|^2}{N_0+P|h_{21}|^2+P|h_{31}|^2}\big),\log_2\big(1+\frac{P|h_{12}|^2}{N_0+P|h_{32}|^2}\big)\big]}$}                      \\ \hline
\hspace{-0.1cm}2\hspace{-0.1cm} & \scriptsize{$\hspace{-0.1cm}\min{\big[\log_2\big(1+\frac{P|h_{21}|^2}{N_0+P|h_{31}|^2}\big),\log_2\big(1+\frac{P|h_{22}|^2}{N_0+P|h_{12}|^2+P|h_{32}|^2}\big)\big]\hspace{-0.1cm}}$}\\ \hline
\hspace{-0.1cm}3\hspace{-0.1cm} & $\min{\big[\log_2\big(1+\frac{P|h_{31}|^2}{N_0}\big),\log_2\big(1+\frac{P|h_{32}|^2}{N_0}\big)\big]}$     \\ \hline
\end{tabular}}
\end{table}

Following that, the second stage is considered, where the delivery of data to the DC is studied. In the second stage, two optimization problems are considered based on the length of the GW buffer. For small-size buffers, the goal is to transmit all the stored data in the transmission slot. In this case, two objective criteria are proposed. The first criterion considers the optimization of the total power expenditure at the GWs. Each GW is assumed to be constrained with a maximum transmission power which helps to save its battery. The second criterion considers power fairness between the GWs. In this case, only the maximum power is considered as the variable for optimization. It is proved that for both objective functions the problem is convex and, hence, convex optimization techniques can be used. The solution is obtained by finding the optimal values of power and the optimal decoding order at the DC. Furthermore, an optimization problem is formulated and proposed for large-size buffers at the GWs. In this case, the goal is to maximize the weighted-sum rate of the GWs. Each GW might have a different length of stored data in its buffer and, hence, it might be necessary to give priority to those GWs with long queues. Giving priority can be mathematically done by increasing the weight of those GWs in the objective function of the optimization problem. The larger the weight of the GW, the more probable that it will have better rate. The problem here is a power allocation problem where the total power is constrained by a maximum value. In this case, the GWs' rates are also constrained by all combinations of upper bounds on the achievable rates. If the survey area changes, the solution has to be updated by solving the relevant problem. As can be seen, this work can be generalized and applied to other WSNs, and not only to seismic acquisition systems.
\begin{table}[]
	\centering
	\caption{Capacity Bounds for the example given in Fig. 2}
	\label{Casestudy1}
	\small{%
		\begin{tabular}{|c|c|c|c|c|c|c|}
			\hline
			\textbf{GP} & \textbf{Rate (max-sum-rate)} & \textbf{Rate (decode-all)}  \\ \hline
			1 & $3.010$  & $\min{(1.640, 0.776)}$  \\ \hline
			2 & $\min{(0.367, 1.335)}$ & $\min{(1.738, 1.335)}$ \\ \hline
			3 & $0.435$ & $\min{(0.372, 0.850)}$  \\ \hline
	\end{tabular}}
\end{table}

\section{Problems Formulation}
The mathematical formulation of the proposed problems is presented in this section, where the first problem considers the sum-rate maximization of the GPs in the first stage. The following two problems consider the second stage where the delivery of data to the DC is discussed.
\subsection{Sum-rate Maximization}
The sum-rate maximization problem is formulated as an integer constrained optimization problem as follows:
\begin{subequations}
\begin{align}
&\max_{F_{ji}} \sum_{j=1}^{K} R_j, \qquad  \text{subject to} \\
&\nonumber \sum_{j \in \Omega}\hspace{-.1cm} F_{ji}R_j \hspace{-.1cm}\leq\hspace{-.1cm}{\log_2\hspace{-.1cm}
	\Bigg(\hspace{-.1cm}{1\hspace{-.1cm}+\hspace{-.1cm}\frac{\sum_{j \in \Omega}F_{ji}P_j|h_{ji}|^2}{N_0\hspace{-.1cm}+\hspace{-.1cm}\sum_{\substack{m=1\\ m\neq j}}^{K}(1\hspace{-.1cm}-\hspace{-.1cm}F_{mi}P_m|h_{mi}|^2)}}}\Bigg),\\
& F_{ji}\in \{0,1\}, \forall i\in \{1,2,\hdots ,N\}, \forall \Omega ,
\end{align}
\end{subequations}
where $R_j$ in (2a) is the actual rate of the $j^{th}$ GP and $\Omega$ in (2b) is an element in the set of all combinations $\binom{K}{b}$, $\forall b\in  \{1,2,\hdots ,K \}$. Each link has an index $F_{ji}$  that takes the value 1 if the $i^{th}$ GW decides to decode the signal of the $j^{th}$ GP, otherwise $F_{ji} = 0$.
The set of constraints in (2b) is a generalized form of the multiple access channel constraints (with multiple destinations), and it gives all the combinations of upper bounds on the achievable rates for all the GWs. The numerator inside the $\log$ in (2b) accounts for the sum of GPs' powers decoded at the $j^{th}$ GW, while the denominator represents the noise power added to the interference from all the GPs not decoded at this GW. Although allowing a GP signal to be decoded at multiple GWs can introduce redundant data, it is essential to maximize the sum-rate of all GPs. This is because decoding a GP's signal at multiple GWs could reduce the interference seen by other GPs' signals and improve the overall sum-rate. The optimization in (2) is performed on the links indices; $F_{ji}$'s. This renders the problem an integer programming problem, where the search space is comprised of all decoding combinations and its size, $\zeta$, is given by the following formula:
\begin{equation}
\zeta = \left [\sum_{i=0}^{K}\binom{K}{i}\right ]^N,
\end{equation}

To give an idea about how rapidly the size of this search space increases, consider the network shown in Fig.~\ref{fig:network example}, where $K=3$ and $N=2$  then from (3), $\zeta = 64$  combinations, while when the parameters are increased to $K=30$  and $N=5$, there are $\zeta =1.427\times 10^{45}$ combinations. Going through all the combinations is a challenging task and consumes huge resources in terms of time and computations. Therefore, it is necessary to look for heuristic and metaheuristic methods to solve this problem. Heuristic algorithms are problem-specific low-complexity search protocols that provide a sub-optimal solution by exploring the search space in a faster way. On the other hand, metaheuristics are high-level algorithms that are applicable on a wide-range of problems. They are generally used to solve complex problems where it is hard to develop a specific algorithm \cite{Boussaid2013}. In Section IV, three sets of metaheuristic algorithms are introduced to solve the problem in (2). Specifically, two variations of particle swarm optimization (PSO), two variations of ant colony optimization (ACO), and the simulated annealing (SA) method are discussed.

\subsection{Min-total Power}
The min-total power considers the case of data delivery from the GWs to the DC with minimal total transmission power. It is formulated as an optimization problem, where the variables of optimization are the individual transmission powers of the GWs, and the decoding order at the DC. The importance of the decoding order is that it affects the individual rate for each GW. The problem is formulated below,
\begin{align}
\nonumber\min_{P_i,\forall i} &P_{\text{Total}}\qquad \text{subject to} \\
\nonumber 0\leq & P_i \leq P_{max} \\
 \sum_{i\in \Omega}Q_i \leq &\log_2 \Big ( 1+\frac{\sum_{i\in \Omega}P_i |g_i|^2}{N_0}\Big ), \forall \Omega ,
\end{align}
where $P_{\text{Total}} = \sum_{i=1}^{N}P_i$ and $g_i$ is the channel gain of the link between the $i^{th}$ GW and the DC. Here, $Q_i$ refers to the normalized data rate in bits/s of the $i^{th}$ GW such that $Q_i = \frac{q_i}{T}$ where $q_i$ is the amount of data stored in the $i^{th}$ GW, in bits, and $T$ is the transmission time, which is a fixed time slot for the GWs. Given the constraints in (4), the GWs powers are optimized, and from the optimal power values, the optimal decoding order can be constructed. The objective function, $P_{\text{Total}}$, is a linear function and is therefore a convex function. The first set of constraints on $P_i, \forall i\in \{1,2,\hdots ,N\}$ are linear constraints too. Thus, each one of them forms a convex set. The upper bounds on the normalized data are of the type $\log(1+ax+by)>c$ where $a$, $b$ and $c$ are constants. Since the $\log$ function is a concave function, these set of constraints represent a convex set. Therefore, the problem is convex, and the global optimal solution can be efficiently obtained using convex optimization techniques such as Newton's method.
The optimal decoding order at the DC for the problem in (4), that will minimize the total power expenditure, is to decode the signals of the GWs based on the channel gain values in descending order. The proof is given in Appendix 1.

\subsection{Min-max Power}
In min-max power, the power fairness between the different GWs is considered. The objective of this problem is to find the decoding order at the DC that will minimize the maximum required transmission power at any of the GWs. The problem is formulated as a minimax problem, where the variables of optimization are the individual transmission powers of the GWs. In the case where the buffers are limited, the GWs must transmit all the stored data before the arrival of the new data. The set of all possible MAC achievable rate combinations form the constraints on the normalized data rate. For practical consideration, each GW is assumed to have a maximum power of $P_{max}$. The problem can be formulated as:
\begin{align}
\nonumber\min_{P_i,\forall i} &\max_{i} P_{i}\qquad \text{subject to} \\
\nonumber 0\leq & P_i \leq P_{max} \\
 \sum_{i\in \Omega}Q_i \leq &\log_2 \Big ( 1+\frac{\sum_{i\in \Omega}P_i |g_i|^2}{N_0}\Big ), \forall \Omega .
\end{align}

Similar to min-total power, the convexity of the min-max power can directly be proved. Once the optimal powers are achieved, the decoding order can be found from the active constraints. A possible scenario is that the DC uses time-sharing between two or more decoding orders in order to achieve the desired rates and at the same time minimize the maximum transmission power. For time-sharing between two decoding order combinations, the sum-rate will be the maximum value and its point will lie on the line connecting between the two corners in the capacity region characterized in (5). To achieve any point on the line, the DC will alternate between the two corners by giving a weight to each decoding order characterized by the corner. This may, sometimes, prove useful to achieve power fairness, by working at a point, where the maximum power of the GWs will be less than the value achieved by working at any of the individual decoding order combinations. A system of linear equations is solved to find the exact percentage of time for each decoding order at which the DC should employ successive decoding, to achieve the desired rate point on the MAC.

\subsection{Max weighted-sum}
 The objective of the max weighted-sum problem is to maximize the weighted sum of the rates of the GWs. Each GW will enjoy a rate that is proportional to its buffer size. The longer the queue of data in the buffer, the larger is the weight in the objective function. A constraint of maximum power is held on the total power of the GWs. Hence, the problem is to allocate the powers for each GW such that the objective function is maximized. A set of constraints, that is comprised of all MAC achievable rate combinations, is also imposed on the objective function. The problem can be formulated as:
\begin{align}
\nonumber\max_{P_i,\forall i}  & \sum_{i=1}^{N} w_i R_{i}\qquad \text{subject to} \\
\nonumber 0\leq & P_{\text{Total}} \leq P_{max} \\
 \sum_{i\in \Omega}R_i \leq &\log_2 \Big ( 1+\frac{\sum_{i\in \Omega}P_i |g_i|^2}{N_0}\Big ), \forall \Omega,
\end{align}
where $w_i$ is the normalized weight of the $i^{th}$ buffer and $\sum_{i=1}^{N}w_i = 1$. The total power $P_{\text{Total}} = \sum_{i=1}^{N}P_i$, is the sum of all powers of the GWs. As can be seen in the objective function, the larger the queue of a buffer of a certain GW, the more weight it will have. In this problem, the queues of the buffers are assumed to be large. This is because the time slot is assumed to be equal in size for all transmission sessions. When the stored data in the buffers is small in size such that all the data can be sent within the time slot, a different problem will appear. The GW that has higher rate will finish sending its data faster and then it will have zero rate. When this happens, the upper bounds on the achievable rates will be altered as the rate of some of the GWs will be zero at some point during the transmission period. This renders the problem very complex to solve. Therefore, the focus here is on the case of long buffers filled with large queues of data, such that all upper bounds on the achievable rates hold throughout the transmission time.

\section{Optimization Algorithms}

In this section, the algorithms to solve the optimization problem in (2) are adopted and conclusion is made based on performance and complexity.

\subsection{Particle Swarm Optimization}

Particle Swarm Optimization (PSO) uses simple velocity and position equations to update the particles positions which correspond to the possible solution \cite{Iqbal2015, Marini2015, RezaeeJordehi2012, Kaur2018, Yan2018, Iqbal2014}. Velocity here represents the rate of change of a particle's position in the search space. Two variations of PSO are used in this work for the problem in (2), namely, angle-modulated particle swarm optimization (AMPSO) and discrete particle swarm optimization (DPSO). These variations are able to solve the binary problem at hand.

\subsubsection{AMPSO}

The AMPSO algorithm has proven efficiency, especially when the dimensionality of the binary vector is large \cite{ Yavuz2016}. The function given in (7), which is derived from angle-modulation theory, is optimized using PSO. It is given as,
\begin{equation}
H(x)=\sin(2\pi (x-a)b \cos(A))+d,
\end{equation}
where $A=2\pi c(x-a) $. $a$, $b$, $c$, $d$ and $x$  are horizontal shift,   maximum frequencies of the sine and cosine functions, vertical shift and a single element from a set of evenly separated intervals based on the number of bits, respectively. The parameters in (7) are substitued into the velocity equation to update their values and obtain the positions of the new parameters. The velocity equation is given below:

\begin{alignat}{2}
\nonumber V_n(t+1)  &= \Phi V_n(t) + C_1R_1 (L_n(t)-S_n(t)\\
&+C_2R_2 (G(t)-S_n(t))),
\end{alignat}
where $\Phi$  is the inertia factor, $L_n(t)$ and $G(t)$  are the local and global best values, respectively. $C_1$  and $C_2$  are the weights given to the local best and global best values, respectively. $R_1$  and $R_2$  are random numbers that are drawn from a uniform distribution between 0 and 1, $U(0,1)$.  $S_n(t)$  is the current optimized parameter, i.e. $a$, $b$, $c$ or $d$. The position update equation is given simply as: $S_n(t+1) = S_n(t) + V_n(t+1)$.


\subsubsection{DPSO}

In DPSO, the indices $F_{ji}$'s in (2) are considered as a matrix of bits, where each row represents the links between the GPs and a specific GW, and each bit (index) is optimized separately. The position of the optimized variable is restricted to 0 or 1. Velocity here is the probability that a bit is 1, where $y$  represents the index of the particle and $z$  represents the index of the dimension. The dimensionality of the bits matrix is $D=N\times K$. The velocity formula is given as \cite{Yang2016}:
\begin{equation}
V_{yz}=V_{yz}+\phi_1 (p_{yz}-x_{yz})+\phi_2 (p_{gz}-x_{yz}),
\end{equation}
where $x_{yz}$  is the position of the $y^{th}$ particle in the $d^{th}$ dimension, $p_{yz}$  is the bit position of the best performance the $y^{th}$ particle has achieved so far. $p_{gz}$  is the bit position for the best performance of all the particles so far (global solution). Obviously, $V_{yz}$ is a real-valued number and to translate it to a probability, a logistical transformation called Sigmoid function is given as:
\begin{equation}
S(V_{yz})=\frac{1}{1+\exp({-V_{yz}})},
\end{equation}

To determine $x_{yz}$ , a random value $\rho$ is generated from a uniform distribution in the range $[0,1]$ . Then, it goes through the following decision rule:
\begin{align}
\nonumber &\text{if} \quad  \rho < S(V_{yz}),  \quad  x_{yz} =1 ,\\	
&\text{else} \quad     x_{yz}=0.
\end{align}

\begin{figure}
	\centering
	\includegraphics[width=9cm]{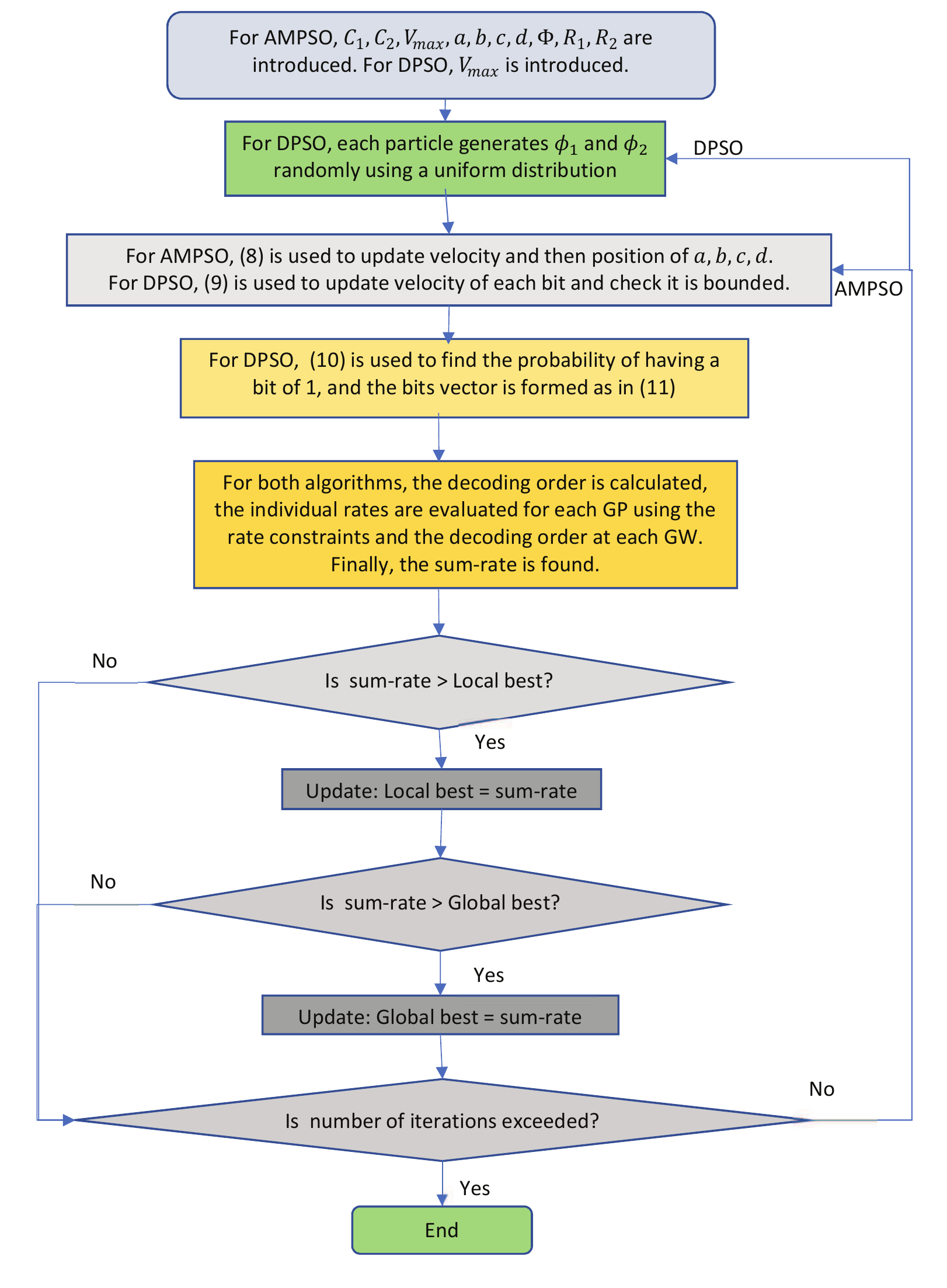}\\
	\caption{AMPSO and DPSO flowchart}
	\label{fig:ampso_dpso}
\end{figure}

A bound is usually set on velocity so that the algorithm keeps exploring the search space. The flowchart in Fig.~\ref{fig:ampso_dpso} illustrates the detailed steps of the AMPSO and DPSO algorithms. It shows the variables initialization stage, followed by the process for each iteration. At a given iteration, each particle updates its potential solution based on the corresponding velocity equations and then updates the position. The local and global best solutions are then updated by comparison with the new solutions offered by the particles.

\subsection{Ant Colony Optimization}

\begin{figure}
	\centering
	\includegraphics[width=7cm]{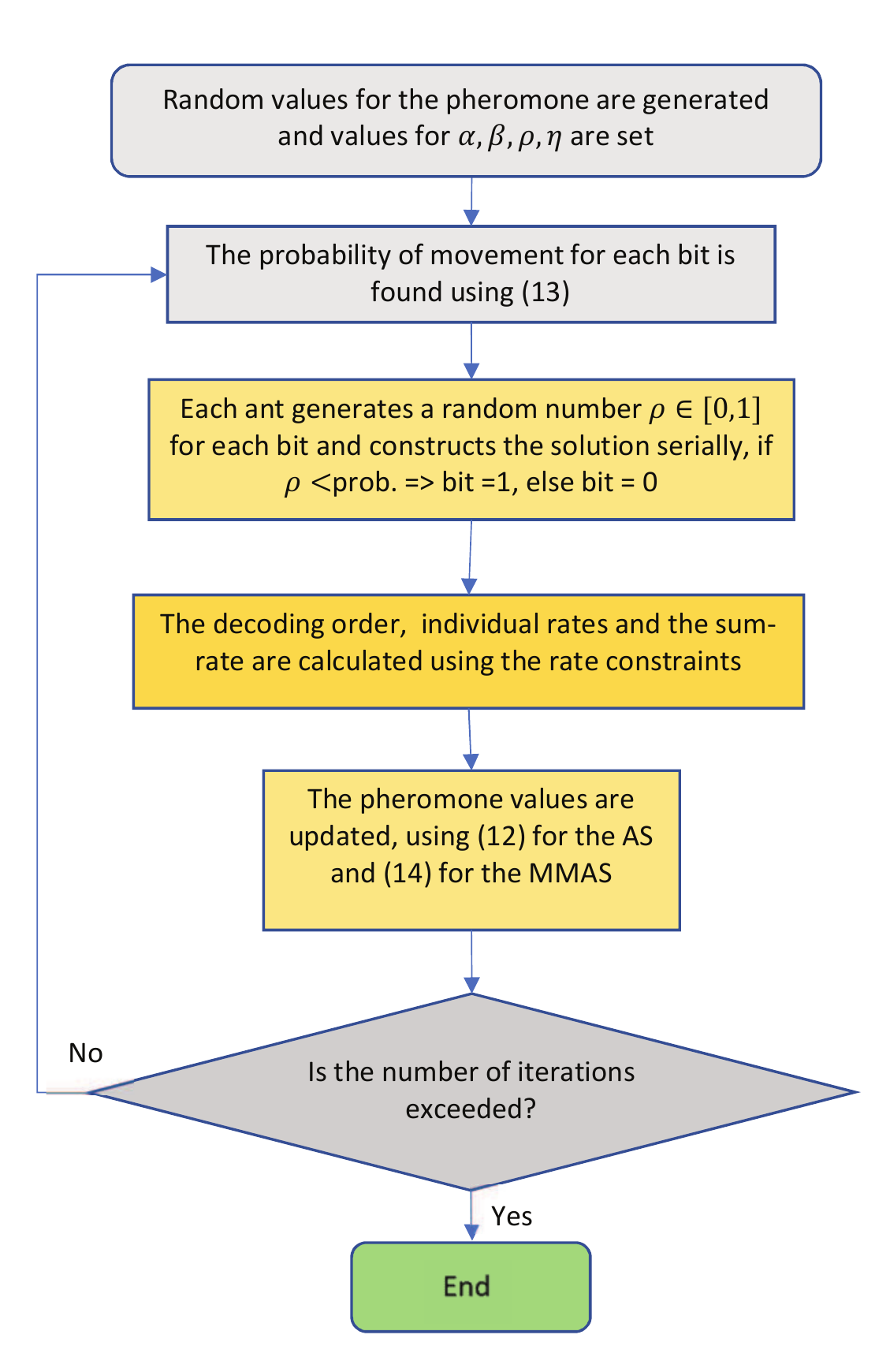}\\
	\caption{A flowchart of the detailed steps of the ACO algorithm}
	\label{fig:aco}
\end{figure}

Ant Colony Optimization (ACO) is based on observations of the behavior of real ants\cite{Dorigo2019,Marzband2016, Zhang2018}. In order to remember the path towards home and guide other ants, while searching randomly for food, they release a chemical substance called pheromone. The pheromone evaporates with time, which allows the ants to explore other parts of the search space and avoid being trapped in a local optimum. If there is more than one path that is taken by ants, the shorter path will hold pheromone longer since ants take less time walking through it. Thus, eventually all ants use this shorter path.
The detailed steps of the general ACO algorithm can be followed in Fig.~\ref{fig:aco}. Firstly, an initialization stage is run by introducing the pheromone and simulation variables. The probability for each binary value is estimated by each ant for all possible paths. Then, ants construct the solution by generating random values and comparing them with the estimated probabilities. Each solution is used to find the sum-rate and update the pheromone.
Varieties of ACO algorithms have been developed in the literature.
\subsubsection{Ant System (AS)}
In AS algorithm, all ants that successfully constructed a solution update the pheromone values at each iteration.
\begin{equation}
\tau_{cd}^{new}=(1-\gamma)\tau_{cd}^{old}+\sum_{r=1}^{m} \Delta \tau_{cd}^{r},
\end{equation}
where $\gamma$  is the evaporation coefficient, $m$ is number of ants, $\tau$ is the pheromone value, $c\in\{0,1\}$  is the index of possible outcomes, $d\in\{1,2,\hdots,D\}$, $D$  is the dimensionality of the system, $D=K\times N$, $\tau_{cd}^{new}$ is the new value of pheromone for a link $d$ with possible outcome $c$ and $\Delta \tau_{cd}^{r}$ is the pheromone deposit, made by the ant $r$, based on its sum-rate value. When constructing a solution, the probability of ant $q$ choosing $c\in\{0,1\}$  for the $d^{th}$ bit is given by:
\begin{equation}
p_{cd}^{q}=\frac{\tau_{cd}^{\alpha}\eta_{cd}^{\beta}}{\sum_{c\in{N(s^P)}}\tau_{cd}^{\alpha}\eta_{cd}^{\beta}},
\end{equation}
where $N(s^P)$ is the set of feasible components, i.e. $\{0,1\}$ and $\eta_{cd}$ is the heuristic information, which depends on the problem at hand. It is usually linked to a priori information about the problem. In our problem, $\eta$ is linked to the channels' gains, as will be shown in Section V. $\alpha$  and $\beta$  represent the weight given to the pheromone and to the heuristic information, respectively.

\subsubsection{Max-Min Ant System (MMAS)}

In MMAS, only the best ant updates the pheromone values. Moreover, the pheromone update is bounded as follows,
\begin{equation}
\tau_{cd}^{new}=[(1-\rho)\tau_{cd}^{old}+ \Delta \tau_{cd}^{r}]_{\tau_{min}}^{\tau_{max}},
\end{equation}
where $\tau_{max}$ is the maximum possible value of the pheromone and $\tau_{min}$ is the minimum possible value of pheromone.
\subsection{Simulated Annealing(SA)}

\begin{figure}
	\centering
	\includegraphics[width=7cm]{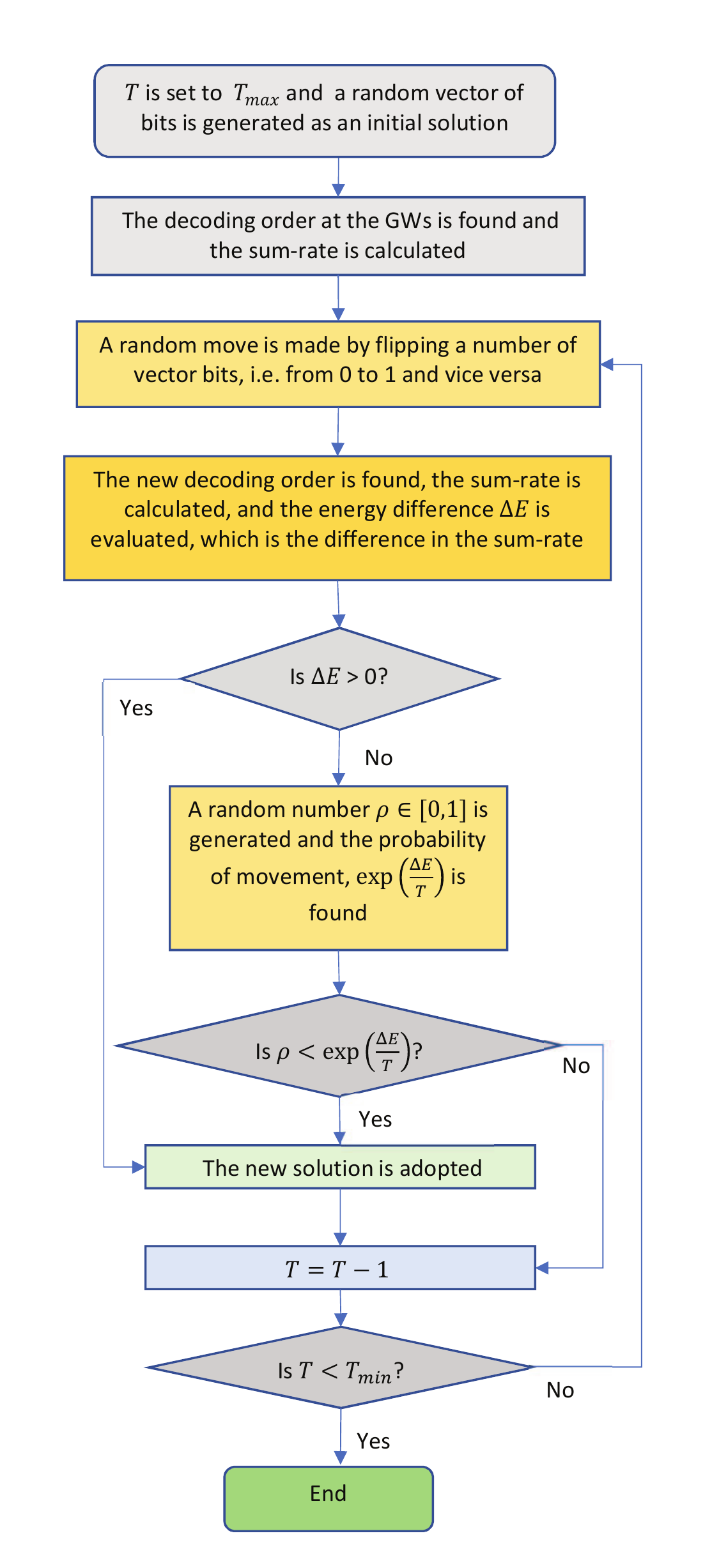}\\
	\caption{A flowchart of the detailed steps of the SA algorithm}
	\label{fig:sa}
\end{figure}

Simulated annealing is based on the metal annealing in material science \cite{Mafarja2017}. It uses restarts to avoid being trapped in a local maximum. It always accepts moves that improve the value of the objective function. If this move does not improve the objective function, then the algorithm may accept the move with some probability. This probability assumes its highest value at the beginning of the algorithm. Therefore, the algorithm is more willing to accept moves that do not improve the objective function at the beginning to explore the search space. As the number of iteration increases, that probability decreases. The willingness of the algorithm to explore new parts of the search space is signified by a temperature degree. Therefore, as it decreases gradually, the algorithm becomes less willing to trade a worse position by its current position. Moreover, the temperature value is equivalent to the number of iterations. This algorithm is illustrated in Fig.~\ref{fig:sa}  where $T$  is the temperature value, $\Delta E$  is the difference in sum-rate between the best solution so far and the next solution.

In Section V, the algorithms are tested and a conclusion is derived about them based on complexity and performance.
\section{Results and Discussion}

\begin{figure}
	\centering
	\includegraphics[width=8cm]{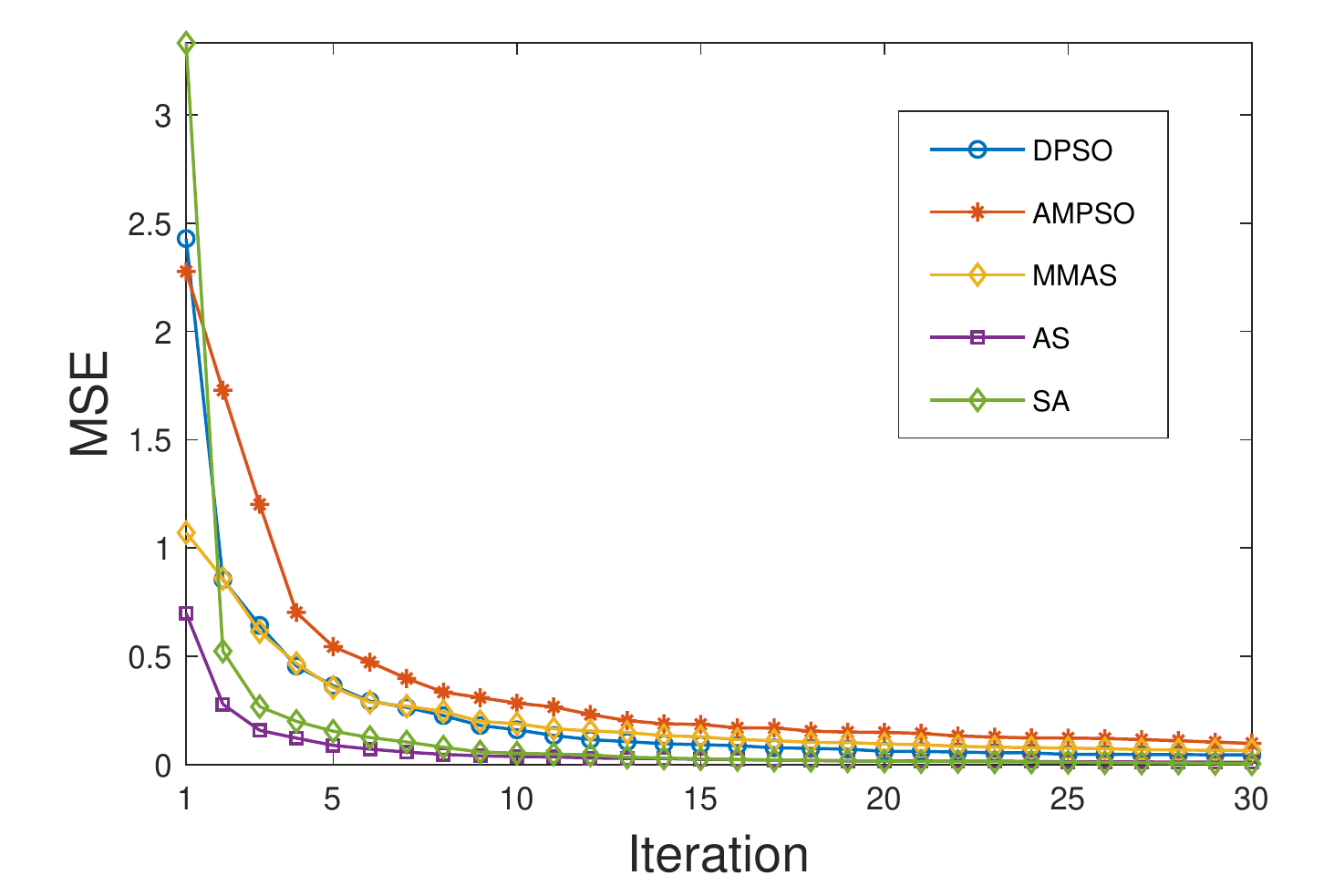}\\
	\caption{MSE versus Iterations for $K=8$  GPs and $N=2$  GWs}
	\label{fig:MSE_Iterations}
\end{figure}

To solve the first optimization problem, a limited computational budget is given to the proposed algorithms in the simulations. The computational budget is represented as the maximum temperature ($T$) for SA, number of particles/ants ($M$) and iterations ($I$) for MMAS/AS/DPSO/AMPSO. The constants used in aforementioned algorithms, $C_1$ ($C_2$), $\Phi$, $V_{max}$ for AMPSO, $V_{max}$ for DPSO, $\alpha$ ($\beta$), $\tau_{max}$  ($=-\tau_{min}$) are set to $1.496$, $0.729$, $4$, $6$, $1$ and $7$, respectively. These values are obtained using extensive simulation and based on \cite{Pampara2005}. In AS, $\eta_{cd}$ equals the channel gain when $c=1$  and equals the average of the other channel gains for the same GW when $c=0$. However, as will be discussed shortly, $\eta$ must be adapted when the problem involves a very large search space. In MMAS, a moderate range of $\tau_{max}$ is found to give acceptable results. Furthermore, a big range might cause convergence stagnation, as the algorithm will keep exploring without benefitting from favourable findings.
The mean square error (MSE) for the algorithms is calculated with respect to the optimal solution. The optimal solution is obtained via exhaustive search (ES) which tries all possible decoding order combinations and finds the one that gives the maximum sum-rate. Fig.~\ref{fig:MSE_Iterations} shows the MSE versus iterations for $K=8$  GPs and $N=2$  GWs. The size of the search space that the ES looks through is $\zeta= 65,536$ combinations. For the algorithms here, a computational budget of $M=30$  and $I=30$ is used. Furthermore, the results are averaged over 100 random channel realizations. As can be seen in Fig.~\ref{fig:MSE_Iterations}, AS and SA achieve the best convergence results. MMAS and DPSO achieve comparable results. AMPSO, however, has a slower convergence and scores the worst results for the final MSE value. It should be noted that in all results, SA algorithm's iterations are modified to match the computational complexity of other algorithms since it does not entail ants/particles. For example, in Fig.~\ref{fig:MSE_Iterations}, SA result is modified by multiplying the number of iterations by $M$, which is 30.
Table \ref{K=8 N=2} shows a comparison between the performance of exhaustive search (ES) and the proposed algorithms for a small network of 8 GPs and 2 GWs. Here, three different levels of computational budget of the proposed algorithms are shown. Comparable results between all the algorithms can be seen, with DPSO achieving higher results for the first two cases and AS achieving better results for the third case where the computational budget is lower.
To describe realistic surveys, number of GPs and GWs must be larger. Therefore, an example of $K=100$ and $N=8$ is considered. The number of decoding order combinations here is beyond the capabilities of any practical computer; $\zeta \simeq 6.67\times10^{240}$  combinations. Therefore, the use of the proposed algorithms with a limited computational budget becomes very crucial in such cases. Two different scenarios for how the GPs transmit data to the GWs are considered. In Scenario 1, the GP is assumed to have a small buffer. In this case, the GW can hold data only up to the next transmission session. On the other hand, Scenario 2 assumes that the GP is equipped with a large buffer to store data for future time transmissions. This gives the GPs the choice to transmit at a given transmission session or not based on its channels' gains. However, sending the stored data later may affect the accuracy of the seismic information that reaches the DC, especially if the channels of those GPs are in a deep fade for several continuous transmission periods.

\subsection*{\textbf{Scenario 1) (Real-time) where all the GPs always send data:}}

\begin{figure}
	\centering
	\includegraphics[width=8cm]{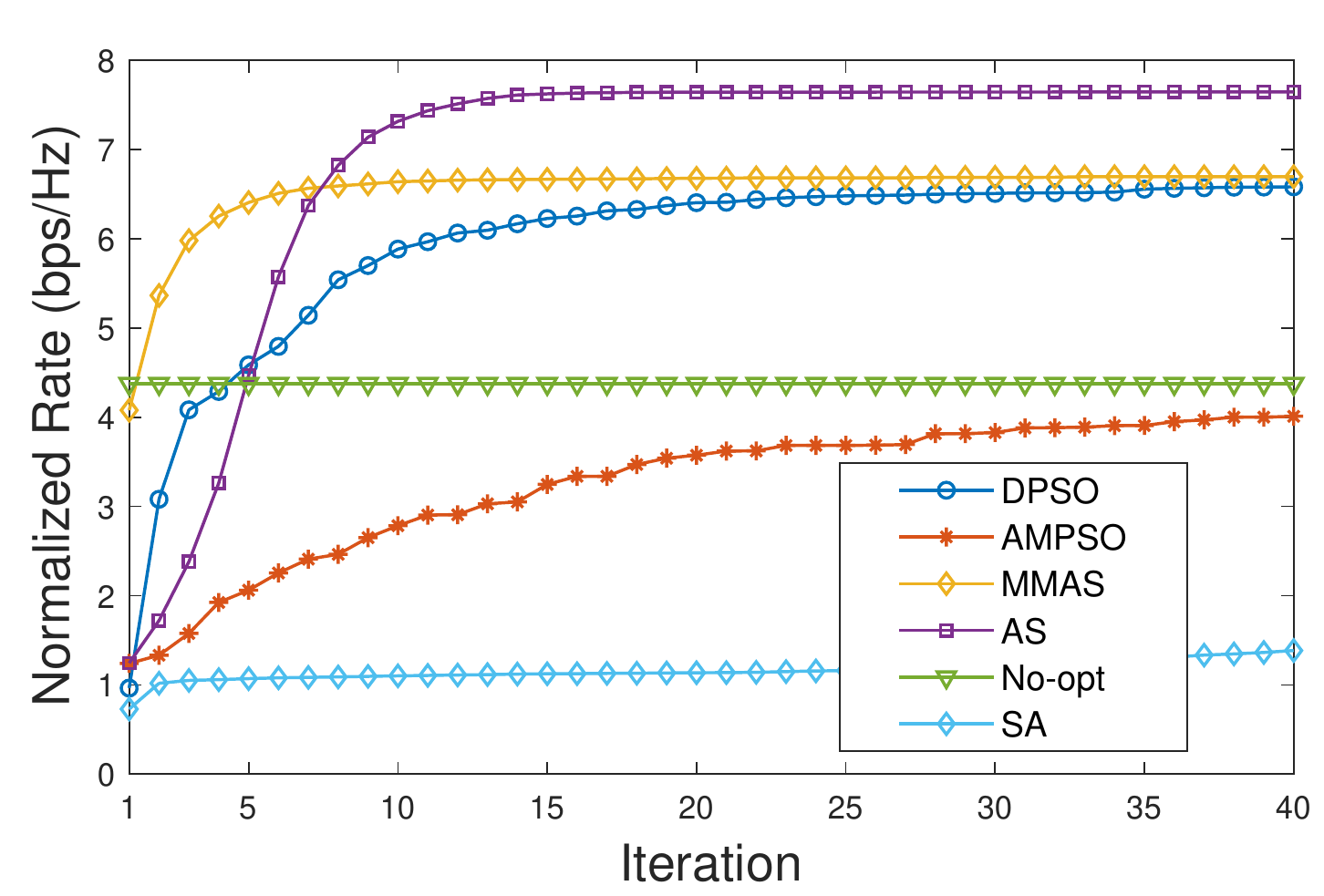}\\
	\caption{Normalized rate versus Iterations for $K=100$ and $N=8$. Scenario 1.}
	\label{fig:Scenario 1}
\end{figure}

In this scenario the powers of all GPs are fixed. Here, it is more beneficial to decode all the signals sent by the GPs due to the existence of interference all the time. Fig.~\ref{fig:Scenario 1} shows the normalized data rate in bps/Hz versus iterations. The results are averaged over 50 random channel realizations. The channels follow Rayleigh fading with $N_0=1\text{mW}$. Here, the used parameters are $M = 250$, $I = 40$ and $T_{max}=M\times I$. As can be seen, AS achieves the best result followed by MMAS, DPSO, AMPSO and finally SA.
\begin{table}[]
	\centering
	\caption{ALGORITHMS PERFORMANCE (bps/Hz) FOR K=8, N=2}
	\label{K=8 N=2}
	\scriptsize{%
		\begin{tabular}{|c|c|c|c|c|c|c|}
			\hline
			\textbf{ES}  & \textbf{MMAS}  & \textbf{AS}  & \textbf{DPSO}  & \textbf{SA}  & \textbf{AMPSO}  & \textbf{Computational budget} \\ \hline
			5.49 & 5.34 & 5.41 & 5.49 & 5.32 & 5.31 & \begin{tabular}[c]{@{}l@{}}$T= 6000$ \\   $M=10$\\   $I=600$\end{tabular} \\ \hline
			5.49 & 5.11 & 5.17 & 5.21 & 4.77 & 4.76 & \begin{tabular}[c]{@{}l@{}}$T= 600$\\   $M=10$\\   $I=60$\end{tabular} \\ \hline
			5.49 & 4.34 & 4.68 & 4.32 & 3.94 & 4.03 & \begin{tabular}[c]{@{}l@{}}$T= 60$\\   $M=1$\\   $I=60$\end{tabular} \\ \hline
	\end{tabular}}
\end{table}

\subsection*{\textbf{Scenario 2) Buffer-Aided GPs:}}
\begin{figure}
	\centering
	\includegraphics[width=8cm]{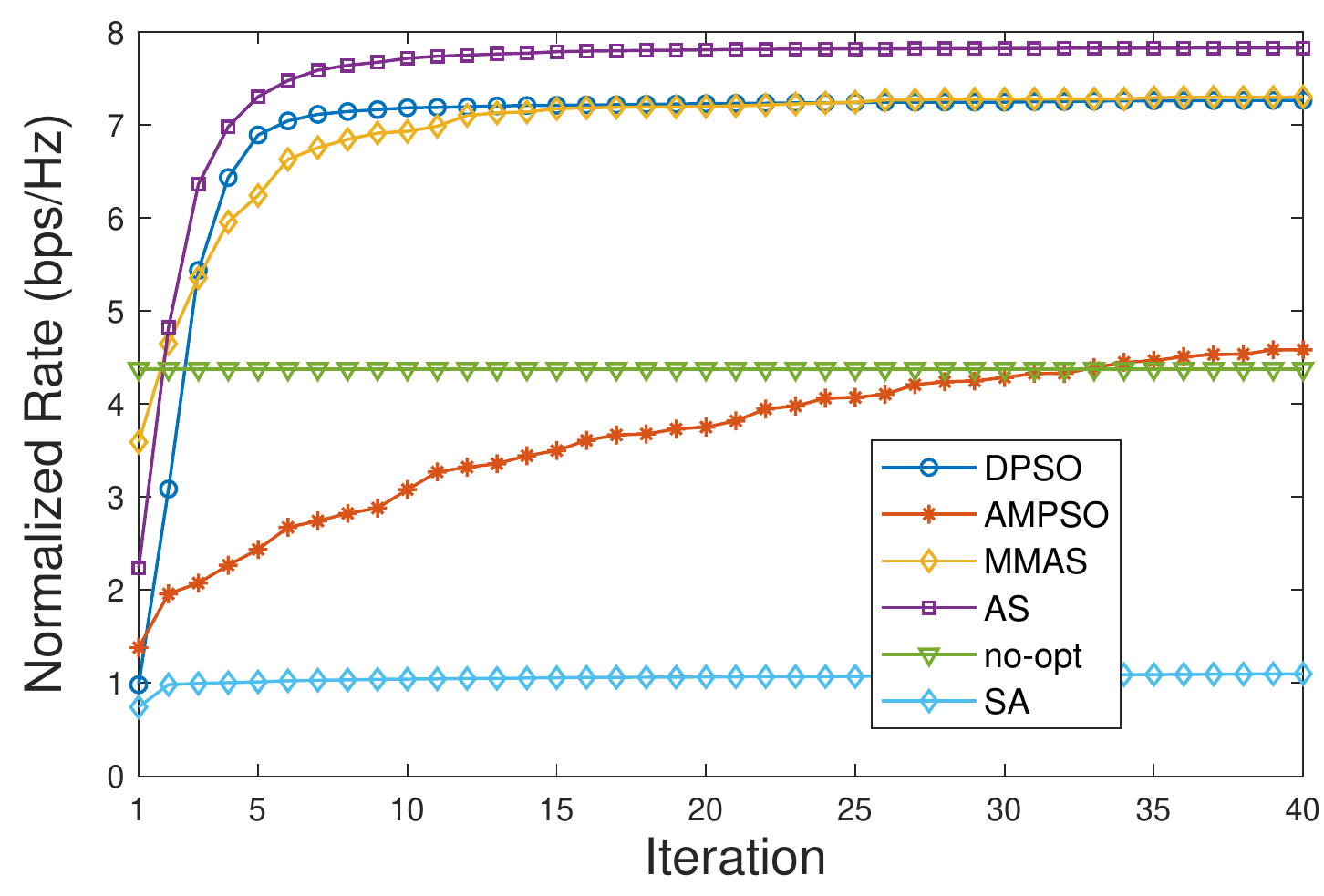}\\
	\caption{Normalized rate versus Iterations for $K=100$ and $N=8$. Scenario 2. }
	\label{fig:Scenario 2}
\end{figure}

All the GPs are assumed to have buffers where they can store the sampled data before transmission. In this scenario the decoding order of the GWs can be optimized so that the sum-rate is even improved over Scenario 1. This is achieved through deactivation of some of the GPs that generally have low channel quality. Fig.~\ref{fig:Scenario 2} shows the normalized rate versus iterations and the results are averaged over 50 random channel realizations. Here, the used parameters are $M = 250$, $I = 40$ and $T_{max} = M\times I$. The best result is found to be achieved by AS, followed by MMAS, DPSO and then AMPSO, while SA displays the worst performance. Table~\ref{scenarios} shows an example comparing the performance of no-optimization and the proposed algorithms for the two scenarios simulating a network of 100 GPs and 8 GWs. The no-optimization case is the special case where the GWs neither cooperate nor share information about the decoding order. It can be seen that Scenario 2 results give comparative advantage in sum-rate over Scenario 1 for most algorithms. Compared to the small network of $K=8$  and $N=2$, it can be seen that SA achieves a lower sum-rate value compared to the no-optimization case.

\begin{table}[]
\centering
\caption{ALGORITHMS PERFORMANCE IN SUM-RATE (bps/Hz) FOR K=100, N=8 }
\label{scenarios}
\footnotesize{%
\begin{tabular}{|c|c|c|}
\hline
\textbf{Algorithms}      & \textbf{Scenario 1} & \textbf{Scenario 2} \\ \hline
No-optimization & 4.33           &  4.33         \\ \hline
DPSO            & 6.32           &  7.51          \\ \hline
AMPSO           & 3.36           &  4.33          \\ \hline
SA              & 1.38           &  1.10          \\ \hline
AS ($\eta$-adapted)  & 7.77      &  7.87          \\ \hline
AS  ($\eta$-not adapted)  & 3.62          &  4.66          \\ \hline
MMAS ($\eta$-adapted)     &  6.92          &  7.17          \\ \hline
MMAS ($\eta$-not adapted) &  1.87          &  2.93          \\ \hline
\end{tabular}}
\end{table}
Also, it is noticed that AS and MMAS achieve lower sum-rate values when their heuristic information is not adapted. This is an interesting observation which indicates the effect of the convergence speed of metaheuristic algorithms on their capability to explore the search space in a limited time. Although AS and MMAS belong to ACO algorithms, which have been proved to always converge, the convergence speed is a problem that is still not fully explored \cite{Krynicki2016, Zhu2007}. To overcome the issue of the algorithm's stagnation, the effects of various parameters such as $\gamma,\alpha,\beta \text{ and }\eta$ are studied. Except for $\eta$, it is found through simulation that the
effects of all parameters on convergence speed are minimal. However, for $\eta$, the choice of the heuristic information will greatly influence the results. This is because $\eta$ biases the search towards regions that are expected to be promising. For Scenario 1, a weight is given to each GW that is proportional to the average of the channels gains associated between all the GPs and this GW. The higher the average for a certain GW, it is more likely to decode signals from GPs. This helps increase the sum-rate because when a GW decodes GPs signals, it will successively eliminate the interference, which increases the total sum-rate. For Scenario 2, it would be useful to turn off GPs which have all their links as weak channels.
The reason is that those GPs have insignificant rates, and they cause interference for other users. Therefore, removing them reduces the unnecessarily limiting interference present in the network, which will help increase the rate of other GPs, while only losing the small rate that belongs to those GPs. To do this, the probability of having an index of 0 for the links associated with these GPs is increased. This makes it more probable that GWs will not decode GPs that have low channel gains. This measure is combined with the Scenario 1 measure, where each GW will have a weight to its links that is proportional to the average of the channel gains. The combined effect directs the search towards regions that have a higher sum-rate and thus increases the algorithm performance with limited budget. Table~\ref{Time} shows the time taken by each algorithm to search for optimal solution for the same scenario considered in Table \ref{scenarios}. It can be shown that the algorithms have a complexity on order of $O(IMNK^3)$. By comparing the time, it is found that SA achieves the best results by taking the least time. Following it, AS and MMAS give comparable results, while DPSO and AMPSO achieve the worst results. However, taking performance and complexity into consideration, it can be concluded that AS achieves the best results.

\begin{figure}
	\centering
	\includegraphics[width=8cm]{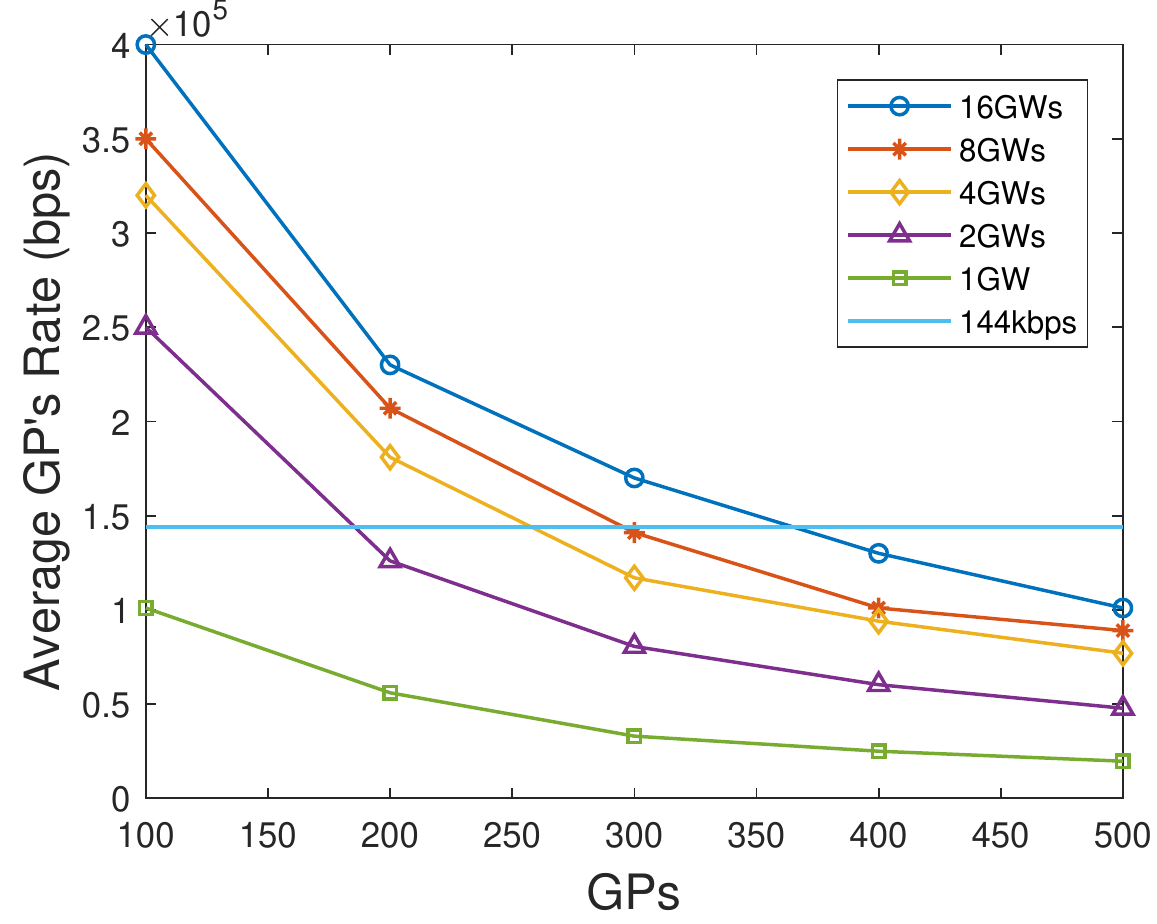}\\
	\caption{Average GP rate versus number of GPs for different GWs using AS. }
	\label{fig:AS_avg_rate}
\end{figure}

\begin{figure}
	\centering
	\includegraphics[width=8cm]{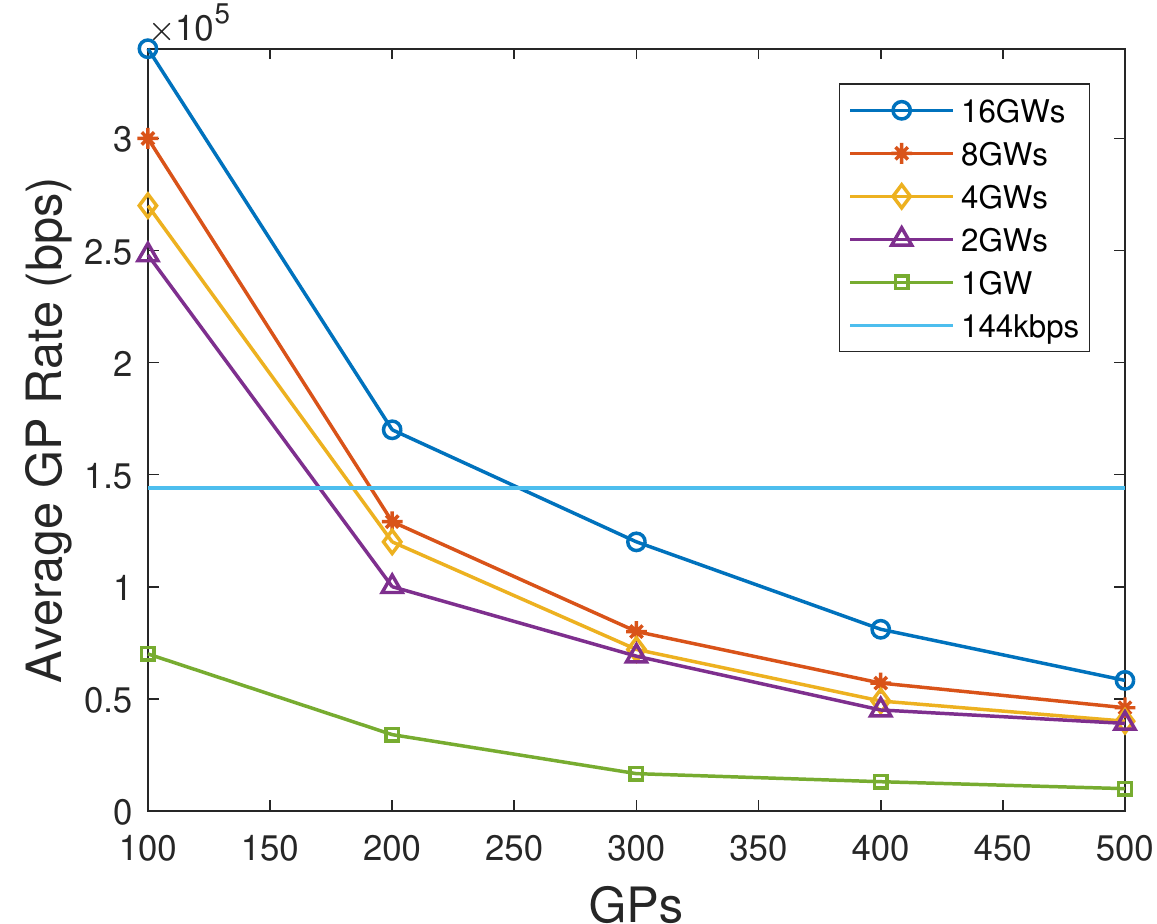}\\
	\caption{Average GP rate versus number of GPs for different GWs using DPSO.}
	\label{fig:DPSO_avg_rate}
\end{figure}

Importantly, our problem formulation cna find the total number of GWs that are required to serve a typical seismic survey. GP's rate(144 Kbps\cite{Iqbal2018}) is used for this purpose. AS and DPSO are used to find the average GP's rate for various number of GPs in Fig.~\ref{fig:AS_avg_rate} and ~\ref{fig:DPSO_avg_rate}, respectively. It is noted that with AS, for example, 16 GWs can support up to around 360 GPs while maintaining the required average GP rate. However, when using DPSO, only up to 250 GPs are served.

In the following, two buffer sizes are considered at the GWs, namely, small-size and large-size buffers.
\begin{table}[]
	\centering
	\caption{ALGORITHMS COMPLEXITY IN TIME}
	\label{Time}
	\footnotesize{%
	\begin{tabular}{|c|c|}
		\hline
		\textbf{Algorithm} & \textbf{Time in seconds}  \\ \hline
		DPSO      & 40            \\ \hline
		AMPSO     & 69.95     \\ \hline
		SA        & 6.67          \\ \hline
		AS        & 13.16       \\ \hline
		MMAS      & 14.5         \\ \hline
	\end{tabular}}
\end{table}

\subsection*{\textit{i}. Small-size Buffer}
The two variations of this problem are solved using the \textit{fmincon} algorithm in MATLAB. The \textit{fmincon} algorithm utilizes a number of techniques to solve convex optimization problems. These techniques include interior-point method and sequential quadratic programming (SQP). These methods are based on approximation of the nonlinear constraints with linear constraints. A case study is considered for eight GWs. Random sets of stored data $Q$, and channel gain values $G$ are generated. The optimal power values for each GW as well as the optimal decoding order at the DC are found via simulation. The total power is also calculated. The same channels and stored data are considered for the two parts of the problem for ease of comparison.

\subsubsection*{\textbf{Case Study}}

 The normalized stored data rates in the buffer of each GW and the channel gain values between the GWs and the DC are given below,

\footnotesize

\begin{align}
\nonumber Q &= [0.996   \quad   1.389 \quad   1.669 \quad 1.219 \quad 1.149 \quad 0.652 \quad 0.913 \quad 1.428],\\
\nonumber G &=[1.095     \quad  0.524 \quad   2.220 \quad 0.967 \quad 1.236 \quad 1.480 \quad 1.837 \quad 0.602],
\end{align}
\normalsize
where $Q$ is in bps/Hz. By solving the min-total power, the individual powers for the GWs are given as,
\begin{align}
\nonumber &P_1 = 13.61 \text{mW}  \quad    P_2 = 5.893 \text{mW}   \quad    P_3 = 94.85 \text{mW} \\
\nonumber &P_4 = 10.02 \text{mW} \quad    P_5= 26.11 \text{mW} \quad
 P_6 = 18.88 \text{mW}\\ \nonumber &P_7 = 29.85 \text{mW} \quad P_8 = 12.20 \text{mW},
\end{align}
and the total power, $P_\text{Total}=211.405 \text{mW}$. The optimal decoding order at the DC is found to be as follows:\\ \small GW\textsubscript{3}$\rightarrow$ GW\textsubscript{7} $\rightarrow$ GW\textsubscript{6} $\rightarrow$ GW\textsubscript{5} $\rightarrow$ GW\textsubscript{1} $\rightarrow$ GW\textsubscript{4} $\rightarrow$ GW\textsubscript{8} $\rightarrow$ GW\textsubscript{2}. \normalsize\\
However, for the min-max power, the solution is found by setting all of the powers of the GWs to be,
\begin{align}
\nonumber &P_i = 46.06 \text{mW},\quad \forall i\in\{1,2,...,8\}
\end{align}
and the total power is, $P_\text{Total}=368.5 \text{mW}$. In this case, the DC uses time-sharing to achieve power fairness. All power values are equal and this is the best possible solution as any decrease of a power value prompts an increase in another to satisfy the achievable rate constraints. The only upper bound constraint that holds with equality is the sum-rate constraint, and this shows that the DC uses time-sharing between a number of decoding order combinations. A system of eight equations and eight variables is solved to find the exact percentage for each decoding order that the DC uses. The decoding order combinations can be chosen randomly but if one or more of the variables is negative, this indicates that the sum-rate point is not in the targeted area. In this case, the decoding order associated with that variable has to be flipped.
The decoding order combinations at the DC are found to be as shown in Table~\ref{timesharing}. Each decoding order combination is shown with the associated percentage of the time. This will minimize the maximum power used for transmission by the GWs.
\begin{table}[]
\centering
\caption{Decoding order combinations at the GWs with time sharing}
\label{timesharing}
\footnotesize{%
\begin{tabular}{|c|c|}
\hline
\textbf{Decoding order combination} & \textbf{Time percentage} \\ \hline
1$\rightarrow$ 2$\rightarrow$ 3$\rightarrow$ 4$\rightarrow$ 5$\rightarrow$ 6$\rightarrow$ 7$\rightarrow$ 8 &   22.38\%         \\ \hline
7$\rightarrow$ 6$\rightarrow$ 5$\rightarrow$ 4$\rightarrow$ 3$\rightarrow$ 2$\rightarrow$ 1$\rightarrow$ 8                            & 3.11\%           \\ \hline
7$\rightarrow$ 8$\rightarrow$ 1$\rightarrow$ 2$\rightarrow$ 3$\rightarrow$ 4$\rightarrow$ 5$\rightarrow$ 6                           & 5.94\%           \\ \hline
6$\rightarrow$ 7$\rightarrow$ 8$\rightarrow$ 1$\rightarrow$ 2$\rightarrow$ 3$\rightarrow$ 4$\rightarrow$ 5                           &  15.63\%          \\ \hline
5$\rightarrow$ 6$\rightarrow$ 7$\rightarrow$ 8$\rightarrow$ 1$\rightarrow$ 2$\rightarrow$ 3$\rightarrow$ 4                            & 14.77\%           \\ \hline
4$\rightarrow$ 5$\rightarrow$ 6$\rightarrow$ 7$\rightarrow$ 8$\rightarrow$ 1$\rightarrow$ 2$\rightarrow$ 3                           &  20.93\%          \\ \hline
3$\rightarrow$ 4$\rightarrow$ 5$\rightarrow$ 6$\rightarrow$ 7$\rightarrow$ 8$\rightarrow$ 1$\rightarrow$ 2                           &  1.14\%          \\ \hline
2$\rightarrow$ 3$\rightarrow$ 4$\rightarrow$ 5$\rightarrow$ 6$\rightarrow$ 7$\rightarrow$ 8$\rightarrow$ 1                           &  16.37\%          \\ \hline
\end{tabular}}
\end{table}
\subsection*{\textit{ii}. Large-size Buffer}
To consider the large-size buffer scenario, an example of a network of eight GWs is considered. The normalized stored data rate in the buffer of each GW and the channel gain value between the GWs and the DC is given below,

\footnotesize

\begin{align}
\nonumber Q &= [87.12   \quad  13.91 \quad   72.25 \quad 98.11 \quad 35.49 \quad 22.04 \quad 71.68 \quad 91.85],\\
\nonumber G &=[0.610  \quad  1.260 \quad   1.920 \quad 1.280 \quad 0.870 \quad 0.560 \quad 1.810 \quad 1.560],
\end{align}\normalsize
where $Q$ is in bps/Hz. $Q$ is generated following the uniform distribution between 30 and 100, i.e. $\in [30,100]$. The problem is solved using the \textit{fmincon} algorithm in MATLAB, where SQP is used. The maximum total power is set to be $P_{max}=5 \text{W}$. From $Q$, the weights for each buffer are calculated to be

\footnotesize
\begin{align}
\nonumber W &=[0.177 \quad 0.028 \quad 0.147 \quad 0.199   \quad 0.072 \quad  0.045 \quad 0.146 \quad 0.187],
\end{align}
\normalsize
where $\sum_{i=1}^{8}W_i = 1$.
The individual powers for the GWs that maximize the objective function are found to be,
\begin{align}
\nonumber P_1 &= 67.88 \text{mW}  \quad    P_2 = 0 \text{W}   \quad    P_3 = 1.694 \text{W} \\ \nonumber P_4 &= 692.8 \text{mW} \quad P_5 = 0 \text{W} \quad P_6 = 0 \text{W}\\ \nonumber P_7 &=  1.331 \text{W}\quad P_8 = 1.214 \text{W},
\end{align}
and the total power, $ P_\text{Total}=5 \text{W}$. The optimal decoding order at the DC is found to be as follows:\\ GW\textsubscript{7} $\rightarrow$ GW\textsubscript{3}  $\rightarrow$ GW\textsubscript{1} $\rightarrow$ GW\textsubscript{8} $\rightarrow$ GW\textsubscript{4}, while GW\textsubscript{2}, GW\textsubscript{5} and GW\textsubscript{6} are turned off.
The normalized data rate for each of the GWs is found to be,
\begin{equation}
\nonumber R = [0.0087 \quad 0 \quad 1.33 \quad  10.146    \quad  0 \quad 0 \quad 0.5047 \quad 1.853],
\end{equation}

In this case, GW\textsubscript{2}, which has the lowest channel gain and lowest stored normalized data rate, is turned off. Moreover, GW\textsubscript{5} and GW\textsubscript{6}, which have relatively low channel gains and stored normalized data rate, are turned off.

\section{Conclusion}
In this paper, the optimization of both stages of a geo-seismic wireless acquisition system were considered. In the first stage, GPs send their sampled data to a few GWs. In this stage, the problem of maximizing the information theoretic bounds on the sum-rate was formulated, and several metaheurisitic algorithms were proposed to solve it. The proposed algorithms are based on concepts from swarm intelligence and material science. These algorithms are angle-modulated particle swarm optimization, discrete particle swarm optimization, ant system, max-min ant system and simulated annealing. Among them, the ant system algorithm proved to be the best in solving the sum-rate maximization problem, considering the performance and computational complexity. Furthermore, it was shown that when the search space gets very large, convergence speed for the ant colony optimization algorithms can be increased by adapting the heuristic information accordingly. In the second stage, which is the data delivery from the GWs to the DC, two problems were explored based on the length of the buffer at the GW. For small-size buffers, an optimization problem to minimize the total power of the GWs and another problem to minimize the maximum power of all the GWs were proposed and solved. For large-size buffers, a problem to maximize the weighted sum of the rates was proposed. Convexity was discussed for the three problems, and simulation results and discussion were provided.

\section*{Appendix 1}
Consider the last constraint on the sum of rates in (4), i.e. $\sum_{i=1}^{N}Q_i \leq \log_2 \Big ( 1+\frac{\sum_{i=1}^{N}P_i |g_i|^2}{N_0}\Big )$. It is obvious that for minimum total power the constraint should hold with equality. Let $A =\sum_{i=1}^{N}Q_i$, where $A$ is a constant. Assume the channel gains $g_i,\forall i\in\{1,...,N\} $ are ordered in a descending order, and assume $P_{N,\text{min}}$ is the minimum possible value for $P_N$, such that,

\begin{equation}
 A = P_1 |g_1|^2 + ... + P_{N-1} |g_{N-1}|^2 + P_{N,\text{min}} |g_{N-1}|^2
\end{equation}

Now, subtract $\Delta_1$ from $P_1$ and add $\Delta_N$ to $P_{N,\text{min}}$ such that the equality still holds, where,
\begin{align}
\nonumber A &= (P_1-\Delta_1)|g_1|^2 + ... + P_{N-1} |g_{N-1}|^2 \\
&+ (P_{N,\text{min}}+\Delta_N) |g_{N}|^2
\end{align}
where $\Delta_N |g_N|^2 = \Delta_1 |g_1|^2$. Therefore, $\Delta_1  = \Delta_N \frac{|g_N|^2}{|g_1|^2}<\Delta_N $.
The total power in this case is,
\begin{align}
 \nonumber P_\text{Total}& = P_1-\Delta_1 +...+P_{N-1}+P_{N,\text{min}}+\Delta_N\\
 \nonumber& = P_1 + ...+P_{N-1}+P_{N,\text{min}}+ (\Delta_N - \Delta_1)\\
 & > P_1+...+P_{N-1}+P_{N,\text{min}},
\end{align}
since $\Delta_N - \Delta_1$ is a positive term. Hence, the minimum value for $P_N$ is used by decoding it last. By repeating the same procedure for $P_{N-1}$ up to $P_{1}$, it is found that the decoding order should follow the decreasing order of the channel gain values.

\section*{Acknowledgment}

This work is supported by the Center for Energy and Geo-Processing (CeGP) at King Fahd University of Petroleum and Minerals (KFUPM) and Georgia Institute of Technology, under grant number GTEC1601.



%

\bibliographystyle{ieeetr}
\bibliography{library}


\end{document}